# Delayed ionization and excitation dynamics in a filament wake channel in dense gas medium

Dmitri A. Romanov[1, 3] and Robert J. Levis[2, 3]

[1]*Department of Physics, Temple University, Philadelphia, PA 19122, USA*

[2]*Department of Chemistry, Temple University, Philadelphia, PA 19122, USA*

[3]*Center for Advanced Photonics Research, College of Science and Technology, Temple University, Philadelphia, PA 19122, USA*

A unified theoretical description is developed for the formation of an ionized filament channel in a dense-gas medium and the evolution of electronic degrees of freedom in this channel in the laser pulse wake, as illustrated on an example of high-pressure argon. During the laser pulse, the emerging free electrons gain energy via inverse Bremsstrahlung on neutral atoms, enabling impact ionization and extensive collisional excitation of the atoms. A kinetic model of these processes produces the radial density distributions in the immediate wake of the laser pulse. After the pulse, the thermalized electron gas drives the system evolution via impact ionization (from the ground and excited states) and collisional excitation of the residual neutral atoms, while the excited atoms are engaged in Penning ionization. The interplay of these three processes determines the electron gas cooling dynamics. The local imbalance of the free-electron and ion densities induces a transient radial electric field, which depends critically on the electron temperature. The evolving radial profiles of the electron, ion, and excited-atom densities, as well as the profiles of electron temperature and induced electric field are obtained by solving the system of diffusion-reaction equations numerically. All these characteristics evolve with two characteristic timescales, and allow for measuring the electronic stage of the wake channel evolution via linear and nonlinear light-scattering experiments.

## 1. Introduction

Laser filamentation in gases has become an area of intense investigation[1-7] due in a large part to the possibilities of controlling physical and chemical processes the filament enables in its wake channel.[8-11] The wake channel evolution includes several processes, which involve electronic and then nuclear degrees of freedom[11,12,13] and proceed at hierarchical timescales. First, the free-electron gas is thermalized on a subpicosecond timescale. Then, the hot electron gas actively exchanges energy with the remaining neutral atoms or molecules in the channel; this stage takes from tens to hundreds of picoseconds, depending on circumstances.[14-16] Then, recombination and possible chemical processes occur on a nanosecond timescale. Finally, the excess energy is transferred to gas dynamics, which leads to formation of a gas density hole on microsecond to millisecond timescale.[11,17,18] Given a typical kilohertz repetition rate of the laser pulses, these latter processes may well create an altered steady-state of the medium.[19,20] Note, however, that the effect of these longer hydrodynamic and heat-conduction processes is predicated on the evolution of the hot electron gas generated by ionization during the filamenting laser pulse.

Strong-field ionization of the medium is an inherent feature of the filamentation process, as the emerging electron gas provides a negative contribution to the refractive index that balances the Kerr self-focusing[4,6] and thus prevents the laser beam collapse. Once generated in the filament, the partially ionized plasma left in the wake of the laser pulse is open to various pump-probe experiments and applications, including backward lasing, four-wave mixing, microwave diagnostics, and Rabi sideband generation.[10,14,21,22]

Various experimental approaches have been used to investigate the evolution of free-electron density in the filament wake, including longitudinal diffraction of a probe beam,[23,24] Rayleigh microwave scattering,[25] and wavefront folding interferometry.[26-28] The entangled dynamics of the electron density and the electron temperature has been also traced indirectly using four-wave mixing in BoxCARS geometry.[14,15,29] In some cases (notably, in argon gas) the electron density and temperature dynamics is also reflected on the dynamic Rabi sidebands related to the manifold of excited states.[22,30]

In typical atmospheric filaments, the free-electron density is rather low, because the ionization rate is stabilized by the mentioned plasma defocusing (so-called intensity clamping[31]), so that only about 0.1% of the gas molecules become ionized. The degree of ionization, however, can be considerably increased in special experimental settings. Thus, in the so-called igniter-heater scheme, the filamenting pulse is accompanied by a second laser pulse of nanosecond duration, which allows one to greatly increase the electron concentration through an avalanche ionization and to make this concentration amenable to control.[32-35] Another approach is to use filamentation of longer (picosecond) pulses at mid-infrared and long-wave infrared wavelengths rather than the widely used 800-nm laser pulses.[36-38] In this case, avalanche ionization becomes significant already during the filamenting pulse and may result in a considerable ion concentration. The third approach to obtaining controllable, high-electron-density wake channels is to use pre-focused laser beams, which have been shown[14,22,23,39] to produce microfilaments of < 30 μm in diameter and <1 cm in length, with free-electron density reaching $10^{18}$-$10^{19}$ cm$^{-3}$. Finally, when filamentation of a standard (800-nm) laser pulse occurs in high-density gases, it also engages secondary impact ionization by the driven electrons. Although not reaching an avalanche scale, this impact ionization plays a significant role, overriding the effect of intensity clamping.[16,40]

In this paper, we address the case of high-density gases. As in this situation collisional processes play a decisive role both during the laser pulse and during the wake-channel evolution, it becomes possible to devise a unified description of these two distinct stages of the medium transformation effected by the filamenting pulse. In our description, we will refer to experimental realization of dense-gas filamentation in argon at the pressure of 60 atm.[16] This high-pressure filamentation regime is important in a number of applications, such as generation of warm dense plasmas,[41] supercontinuum generation,[42-44] and efficient ultrahigh harmonic generation.[45,46]

One major difference from the atmospheric-pressure gases[47] is that in a high-pressure gas the electrons released by strong-field ionization will undergo multiple collisions with neighboring neutral atoms over the duration of the laser pulse, as they are forcefully driven by the oscillating laser field. In this situation, collisional excitation of the constituent atoms or molecules is as important as their impact ionization and may often lead to a preponderance of excited atoms over ionized atoms at the end of the filamenting pulse.[40]

After the laser pulse, the combination of the large concentration of excited atoms and high gas density makes an additional phenomenon actively engaged in the wake channel evolution: Penning ionization.[16] This ionization mechanism involves interaction of two excited atoms, which results in de-excitation of one of them and ionization of the other, with the emerging free electron receiving considerable kinetic energy. While in typical gas-discharge regime the characteristic times of Penning ionization lie in the microsecond range,[50] the conditions in the dense-gas filament wake channel shift this characteristic time into the subnanosecond range and make this process a major actor in the channel evolution.[16]

We will trace the dense-gas medium transformation starting with the arrival of an intense laser pulse, through the formation of inhomogeneous excited mixture of free electrons, ions, and excited neutral atoms. The subsequent evolution of the electronic degrees of freedom in the pulse wake results observable outcomes including electron, ion, and excited-atom density distributions as functions of time, as well as the distributions of electron temperature. There is also a considerable radial electric field induced by the local mismatch of the electron and ion densities.

The paper is organized as follows. In Section 2, we develop a unified description of dense-gas medium excitation during an intense laser pulse (Subsection 2.1) and evolution of the electronic degrees of freedom in the pulse wake (Subsection 2.2), both processes being largely affected by electron collisions with neutral atoms. In Section 3, we develop simplifying approximations to the wake-channel equations and present numerical solutions for the evolution of the densities, electron temperature, and the radial electric field distributions. We then draw conclusions and briefly discuss the implications of our findings in Section 4.

**2. The model description**

Enacted by an intense, femtosecond laser pulse, the transformation of a dense-gas medium comprises two distinct stages. First, during the pulse, build-up of the ionized atoms, excited neutral atoms, and non-equilibrium energetic electrons occurs. Then, when the pulse is over, this system evolves, forming the filament wake channel. Thus, the outcome of the first stage sets the initial conditions for the second stage.

*2.1 Ionization and excitation during the laser pulse*

During the femtosecond laser pulse, the kinetics of the emerging electron gas is essentially local and consists of three major processes: (i) energy gain via inverse Bremsstrahlung; (ii) impact excitation of neutral atoms, which is associated with energy loss by the free electrons; and (iii) impact ionization, which is associated with both the energy loss by free electrons and the generation of new free electrons.

As the number of free electrons is not conserved, we trace this kinetics not in terms of the distribution function but rather in terms of the local density of occupied energy states, $n(\varepsilon,\mathbf{r},t)$, where $\varepsilon$ is the electron-state energy, $\mathbf{r}$ is the position vector, and $t$ is time, so that the instantaneous local electron number density is $n(\mathbf{r},t) = \int_0^\infty d\varepsilon\, n(\varepsilon,\mathbf{r},t)$ and the istantaneous local energy density deposited in the electron gas is $E(\mathbf{r},t) = \int_0^\infty d\varepsilon\, \varepsilon\, n(\varepsilon,\mathbf{r},t)$. The linearly polarized laser field exerts a force on a free electron, $\mathbf{F}(\mathbf{r},t) = \hat{\mathbf{e}} e E_0 s(\mathbf{r},t)\cos(\omega t)$, where $\hat{\mathbf{e}}$ is the laser polarization vector, $E_0$ is the laser field amplitude, and $s(\mathbf{r},t)$ is the dimensionless pulse envelope function. The laser carrier frequency $\omega$ is assumed to be greater than the elastic scattering rate of the electrons with neutral atoms, which in turn is greater than the inverse of the pulse duration, $T$. Given these approximations, the laser-cycle averaged equation for the evolution of $n(\varepsilon,\mathbf{r},t)$ is obtained from Boltzmann's kinetic equation in the form,[40]

$$\frac{\partial n}{\partial t} = \frac{4}{3}\sqrt{\frac{2}{m}}n_0\sigma_0 U_p s^2(t)\frac{\partial}{\partial \varepsilon}\left[\sqrt{\varepsilon}\left(\varepsilon\frac{\partial n}{\partial \varepsilon} - \frac{1}{2}n\right)\right] + \left.\frac{\partial n}{\partial t}\right|_{ex} + \left.\frac{\partial n}{\partial t}\right|_{ion} + g(\varepsilon)W(t) \qquad (1)$$

where $U_p = (eE_0)^2/(4m\omega^2)$ is the ponderomotive potential, $W(t)$ is the rate of strong-field ionization by the laser pulse, and it is taken into account that in the energy range of interest the transport elastic scattering cross-section is approximately constant, $\sigma_{tr}(\varepsilon) \approx \sigma_0 = 10^{-15}$ cm$^2$.[49]

In the right-hand side of Eq. (1) the first term is of Fokker-Plank-type describing the inverse Bremsstrahlung process due to electron elastic scattering on neutral atoms. This process can be seen as an effective diffusion of the electron density along the energy coordinate, with a time-dependent effective diffusion coefficient.

The second term in the right-hand side describes the process of collisional excitation, when a free electron promotes a neutral atom to an excited state with the excitation energy, $\varepsilon_{ex}$, and loses an equivalent amount of kinetic energy: $(\partial n/\partial t)|_{ex} = -\nu_{ex}(\varepsilon)n(\varepsilon,t)\Theta(\varepsilon-\varepsilon_{ex}) + \nu_{ex}(\varepsilon+\varepsilon_{ex})n(\varepsilon+\varepsilon_{ex},t)$, where the two terms correspond to the loss and gain in the electron population of the energy state $\varepsilon$. The collisional excitation rate, $\nu_{ex}(\varepsilon)$, is determined by the concentration of the neutral atoms, $n_0$, the electron velocity, $\sqrt{2\varepsilon/m}$, and the energy-dependent total excitation cross-section, $\sigma_{ex}(\varepsilon)$, in the standard way: $\nu_{ex}(\varepsilon) = n_0\sigma_{ex}(\varepsilon)\sqrt{2\varepsilon/m}$. We restrict our model with one representative excited state having the excitation energy of $\varepsilon_{ex} = 11.8$ eV, use for $\sigma_{ex}(\varepsilon)$ the semi-empirical formula of Ref.[50], and obtain

$$\nu_{ex}(\varepsilon) = 1.40\,\pi n_0 a_B^2 \sqrt{\frac{2\varepsilon}{m}} \left(\frac{Ry}{\varepsilon_{ex}}\right)^2 \left(\frac{\varepsilon_{ex}}{\varepsilon}\right)^{0.75} \left(1-\frac{\varepsilon_{ex}}{\varepsilon}\right)^2 \Theta(\varepsilon-\varepsilon_{ex}) \qquad (2)$$

As a function of energy, $v_{ex}(\varepsilon)$ has a threshold at $\varepsilon = \varepsilon_{ex}$, and rises according to a power law in the vicinity of this threshold.

The third term in the right-hand side of Eq. (1), describes the effects of impact ionization processes and is constructed in a way similar to the second term:

$$\left.\frac{\partial n}{\partial t}\right|_{ion} = -v_{ion}(\varepsilon) n(\varepsilon, \mathbf{r}, t) \Theta(\varepsilon - \varepsilon_{ion}) + v_{ion}(\varepsilon + \varepsilon_{ion}) n(\varepsilon + \varepsilon_{ion}, \mathbf{r}, t) + g(\varepsilon) \int_{\varepsilon_{ion}}^{\infty} d\varepsilon \, v_{ion}(\varepsilon) n(\varepsilon, \mathbf{r}, t) \quad (3)$$

where $v_{ion}(\varepsilon) = n_0 \sigma_{ion}(\varepsilon) \sqrt{2\varepsilon/m}$ is the impact ionization rate, $\varepsilon_{ion}$ is the ionization energy, and $\sigma_{ion}(\varepsilon)$ is the energy-dependent total ionization cross-section, for which we use the celebrated semi-empirical Lotz formula.[51] Although generally the cross-section is given by the sum of subshell contributions, for Ar and for the electron energies in question it is sufficient to consider only the uppermost shell. As a result,

$$v_{ion}(\varepsilon) = 4.896 \cdot \pi n_0 a_B^2 \sqrt{\frac{2\varepsilon}{m}} \frac{Ry^2}{\varepsilon_{ion}\varepsilon} \left(1 - b \cdot \exp\left(-c\frac{\varepsilon - \varepsilon_{ion}}{\varepsilon_{ion}}\right)\right) \ln\left(\frac{\varepsilon}{\varepsilon_{ion}}\right) \Theta(\varepsilon - \varepsilon_{ion}) \quad (4)$$

where $b = 0.62$ and $c = 0.40$ are the empirical constants, and the ionization energy is $\varepsilon_{ion} = 15.76$ eV. The function $v_{ion}(\varepsilon)$ has a threshold at $\varepsilon_{ion}$, and then rises sharply throughout the pertinent energy range. The last term in the right-hand side of Eq. (3) accounts for new free electrons emerging from the impact ionization. These secondary electrons emerge with low kinetic energy, but upon the first cycle of laser-field acceleration and elastic scattering they acquire some initial energy distribution, which is modeled by an auxiliary function, $g(\varepsilon)$. The specific form of this function is of no particular importance; we have chosen $g(\varepsilon) = C_{m\mu}\varepsilon^m \left(1 - \tanh\left(\mu(\varepsilon - U_p)\right)\right)$, where $m$ and $\mu$ are adjustable parameters and $C_{m\mu}$ is the normalization constant.

The local density of the excited atoms, $n_{ex}(\mathbf{r},t)$, grows by the acts of collisional excitation, $dn_{ex}/dt = \int_{\varepsilon_{ex}}^{\infty} d\varepsilon\, \nu_{ex}(\varepsilon) n(\varepsilon, \mathbf{r}, t)$. Likewise, the local ion density, $n_{ion}(\mathbf{r},t)$, grows in step with the total number of free electrons: $dn_{ion}/dt = \int_{\varepsilon_{ion}}^{\infty} d\varepsilon\, \nu_{ion}(\varepsilon) n(\varepsilon, \mathbf{r}, t)$. The outcome of the competition between impact ionization and collisional excitation depends on the relative values and functional dependencies of $\nu_{ex}(\varepsilon)$ and $\nu_{ion}(\varepsilon)$. The function $\nu_{ion}(\varepsilon)$ determined by Eq. (4) rises faster and eventually overcomes the function $\nu_{ex}(\varepsilon)$ determined by Eq. (2). However, $\nu_{ex}(\varepsilon)$ has an earlier onset, and thus there is a window of electron energies in which the collisional excitation is preferable to the impact ionization, as seen in Fig. 1, where the difference between $\nu_{ex}(\varepsilon)$ and $\nu_{ion}(\varepsilon)$ is referred to $\nu_0 = \pi n_0 a_B^2 (Ry/\varepsilon_{ion})^2 \sqrt{2\varepsilon_{ion}/m}$ and is shown as a function of $\varepsilon/\varepsilon_{ion}$. As the energy-gaining electron gas has to cross this window, a relatively large concentration of excited atoms may be expected by the end of the pulse.

Solving Eq. (1), one obtains $n(\mathbf{r},t)$, $n_{ex}(\mathbf{r},t)$, and $n_{ion}(\mathbf{r},t)$. These functions assume some final spatial form at the end of the laser pulse, $t = t_f$, and these spatial distributions serve as the initial conditions for the forthcoming wake-channel evolution. Likewise, the final form of the energy density distribution, $E(\mathbf{r}, t_f)$, determines the electron temperature profile upon thermalization in the pulse wake, $T(\mathbf{r}) = (2/3)\langle \varepsilon \rangle = (2/3) E(\mathbf{r}, t_f) / n(\mathbf{r}, t_f)$.

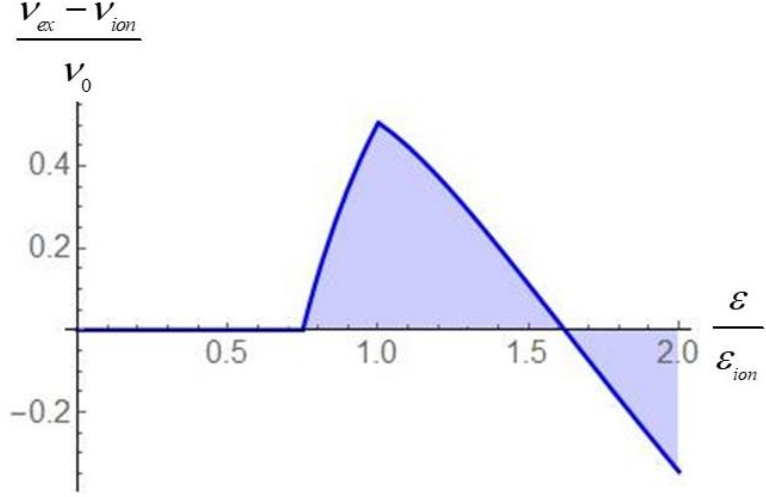

**Figure 1**. The difference between the collisional excitation rate and the impact ionization rate as a function of the electron energy, $\nu_0 = \pi n_0 a_B^2 \left( Ry/\varepsilon_{ion} \right)^2 \sqrt{2\varepsilon_{ion}/m}$.

## 2.2 *Evolution of electronic degrees of freedom in the wake channel*

During the subnanosecond period of the energy redistribution among the electronic degrees of freedom in the filament wake channel, the motion of ions and neutral atoms on the scale of the wake channel diameter can be safely neglected, as well as the changes in the temperature of these heavy particles. Thus, the instantaneous state of the channel is determined by the following distributions: the local free-electron density, $n(\mathbf{r},t)$, the local electron temperature, $T(\mathbf{r},t)$, the ensemble-average (hydrodynamic) velocity of the electron gas, $\mathbf{u}(\mathbf{r},t)$, the ion density, $n_i(\mathbf{r},t)$, and the densities of the neutral atoms in the excited state, $n_{ex}(\mathbf{r},t)$, and in the ground state, $n_{gs}(\mathbf{r},t) = n_0 - n_{ex}(\mathbf{r},t) - n_i(\mathbf{r},t)$, where $n_0$ is the initial homogeneous atomic density. These functions are related among themselves via the processes of impact excitation and impact ionization, and the equations describing the concerted evolution of these functions results from averaging the kinetic equation with the coordinate part taken into account.

The electron density evolution is described by the continuity-generation equation, $\partial n/\partial t + \nabla(n\mathbf{u}) = G_{gs}(n,n_{gs},T) + G_{ex}(n,n_{ex},T) + G_P(n_{ex})$, where $G_{gs}(n,n_{gs},T)$ and $G_{ex}(n,n_{ex},T)$ are the rates of free-electron generation from the ground-state neutral atoms and the excited-state neutral atoms, respectively, via impact ionization by the hot electrons, and $G_P(n_{ex})$ is the rate of the Penning ionization. The two impact-ionization rates are proportional to the density of the incident electrons, $n$, and to the respective densities of the ground-state neutral atoms, $n_{gs}$, and the excited-state neutral atoms, $n_{ex}$, so that $G_{gs}(n,n_{gs},T) = \gamma_{ig}(T) n \, n_{gs}$ and $G_{ex}(n,n_{ex},T) = \gamma_{ie}(T) n \, n_{ex}$, where $\gamma_{ig}(T)$ and $\gamma_{ie}(T)$ are the rate coefficients that depend on the electron temperature. In contrast, the rate of Penning ionization is quadratic in the excited-atom density and independent of the electron temperature: $G_P(n_{ex}) = \gamma_P \, n_{ex}^2$. The evolution of $n_i(\mathbf{r},t)$, $n_{gs}(\mathbf{r},t)$, and $n_{ex}(\mathbf{r},t)$ is determined by the same ionization processes and additionally by the process of impact excitation, whose rate, $P_{ex} = P_{ex}(n,n_{gs},n_{ex},T)$, is likewise the product of the free-electron density, the ground-state atom density, and the impact-excitation coefficient: $P_{ex} = \nu_{ex}(T) n \, n_{gs}$. Having excluded $n_{gs}(\mathbf{r},t)$, the set of density-evolution equations for $n(\mathbf{r},t)$, $n_i(\mathbf{r},t)$, and $n_{ex}(\mathbf{r},t)$ reads:

$$\frac{\partial n}{\partial t} + \nabla(n\mathbf{u}) = n\left(\gamma_{ig}(T)(n_0 - n_i) + (\gamma_{ie}(T) - \gamma_{ig}(T))n_{ex}\right) + \gamma_P n_{ex}^2$$

$$\frac{\partial n_i}{\partial t} = n\left(\gamma_{ig}(T)(n_0 - n_i) + (\gamma_{ie}(T) - \gamma_{ig}(T))n_{ex}\right) + \gamma_P n_{ex}^2 \qquad (5)$$

$$\frac{\partial n_{ex}}{\partial t} = n\left(\nu_{ex}(T)(n_0 - n_i) - (\gamma_{ie}(T) + \nu_{ex}(T))n_{ex}\right) - \gamma_P n_{ex}^2$$

In turn, $\mathbf{u}(\mathbf{r},t)$ and $T(\mathbf{r},t)$ are determined by the gas-dynamic momentum equation and the energy-balance equation. When the ionization and excitation processes are taken into account, these equations read:

$$n\left(\frac{\partial \mathbf{u}}{\partial t} + (\mathbf{u}\nabla)\mathbf{u}\right) = -\frac{1}{m}\nabla(nT) + \frac{e}{m}\mathbf{E}n - \mathbf{u}\left(G_{gs}(n,n_{gs},T) + G_{ex}(n,n_{ex},T) + G_P(n_{ex})\right) - \nu n\mathbf{u}$$

$$n\left(\frac{\partial}{\partial t} + (\mathbf{u}\nabla)\right)\left(\frac{3}{2}T\right) + nT(\nabla\mathbf{u}) - \nabla(\kappa\nabla T) = \qquad (6)$$

$$-G_{gs}\varepsilon_{ion} - G_{ex}(\varepsilon_{ion} - \varepsilon_{ex}) - P_{ex}\varepsilon_{ex} + G_P(2\varepsilon_{ex} - \varepsilon_{ion}) - (G_{gs} + G_{ex} + G_P)\left(\frac{3}{2}T - \frac{mu^2}{2}\right) - \nu n\frac{mu^2}{2}$$

where $\nu = \nu(n,n_i,n_{gs},n_{ex},T)$ is the total elastic collision rate of electrons on ions and neutral atoms. In the right-hand side of the first equation in Eq. (6), the first term is actually the pressure gradient, the second, the volume force, the third signifies that the free electrons are generated with zero velocity, and the last term represents the effective friction of the electron gas against ions and neutral atoms (the viscose forces are neglected). In the second equation in Eq. (6), $\kappa = \kappa(n,n_i,n_{gs},n_{ex},T)$ is the coefficient of thermal conductivity in the electron gas. The second line accounts for the energy spent by the electron gas on the impact ionization and collisional excitation, as well as for the energy gain from Penning ionization. Specifically, one single act of impact ionization of a ground-state atom takes $\varepsilon_{ion}$ of the electron-gas energy, an act of impact ionization of an excited atom takes only $\varepsilon_{ion} - \varepsilon_{ex}$ of energy, and an act of Penning ionization adds $2\varepsilon_{ex} - \varepsilon_{ion}$ to the energy pool. Finally, the local electric field $\mathbf{E}(\mathbf{r},t)$, which acts in the right-hand side of the first equation in Eq. (6) is linked to the electron and ion densities via the Poisson equation, $\nabla \mathbf{E} = 4\pi e (n(\mathbf{r},t) - n_i(\mathbf{r},t))$.

In what follows, we use the phenomenological relation for the thermal conductivity coefficient, $\kappa = (3nT)/(2m\nu)$, and we neglect terms that are quadratic in $u$, based on the fact that $(mu^2/2) \ll T$. Then, in the cylindrical coordinates associated with the filament channel, the total system of equations takes the following form:

$$\frac{\partial n}{\partial t} + \frac{1}{r}\frac{\partial}{\partial r}(rnu) = n\left(\gamma_{ig}(n_0 - n_i) + (\gamma_{ie} - \gamma_{ig})n_{ex}\right) + \gamma_P n_{ex}^2 \tag{7}$$

$$\frac{\partial u}{\partial t} = -\frac{1}{mn}\frac{\partial}{\partial r}(nT) + \frac{e}{m}E - u\left(\gamma_{ig}(n_0 - n_i) + (\gamma_{ie} - \gamma_{ig})n_{ex} + \nu + \gamma_P \frac{n_{ex}^2}{n}\right) \tag{8}$$

$$\frac{\partial T}{\partial t} + u\frac{\partial T}{\partial r} + \frac{2}{3}\frac{T}{r}\frac{\partial}{\partial r}(ru) - \frac{1}{mnr}\frac{\partial}{\partial r}\left(r\frac{nT}{\nu}\frac{\partial T}{\partial r}\right) =$$
$$-\left(\gamma_{ig}\left(T + \frac{2}{3}\varepsilon_{ion}\right) + \frac{2}{3}\nu_{ex}\varepsilon_{ex}\right)(n_0 - n_i) - \tag{9}$$
$$\left((\gamma_{ie} - \gamma_{ig})\left(T + \frac{2}{3}\varepsilon_{ion}\right) - \frac{2}{3}(\nu_{ex} + \gamma_{ie})\varepsilon_{ex}\right)n_{ex} - \gamma_P \frac{n_{ex}^2}{n}\left(T + \frac{2}{3}(\varepsilon_{ion} - 2\varepsilon_{ex})\right)$$

$$\frac{1}{r}\frac{\partial}{\partial r}(rE) = -4\pi e(n - n_i) \tag{10}$$

$$\frac{\partial n_i}{\partial t} = n\left(\gamma_{ig}(n_0 - n_i) + (\gamma_{ie} - \gamma_{ig})n_{ex}\right) + \gamma_P n_{ex}^2 \tag{11}$$

$$\frac{\partial n_{ex}}{\partial t} = n\left(\nu_{ex}(n_0 - n_i) - (\gamma_{ie} + \nu_{ex})n_{ex}\right) + \gamma_P n_{ex}^2 \tag{12}$$

In these equations, the temperature-dependent coefficients $\gamma_{ig}(T)$, $\gamma_{ie}(T)$, and $\nu_{ex}(T)$ are obtained by averaging the rates given by Eq. (4) and Eq. (2) over the thermal distribution of free electrons.

It is convenient to cast Eqs. (7)-(12) in a dimensionless form, using the ionization potential, $\varepsilon_{ion}$, as the characteristic energy scale. The dimensionless temperature is thus determined as $\tilde{T} = T/\varepsilon_{ion}$, and the dimensionless excitation energy as $\tilde{\varepsilon}_{ex} = \varepsilon_{ex}/\varepsilon_{ion}$. Then, the thermal averaging of Eq. (4) results in $\gamma_{ig} = \Gamma_g \tilde{\gamma}_{ig}(\tilde{T})$, where $\Gamma_g = 9.79\pi^2 (Ry/\hbar) a_B^{\,3} (2Ry/(\pi I_p))^{3/2}$ and

$$\tilde{\gamma}_{ig} = -\frac{1}{\sqrt{\tilde{T}}}\left(Ei\left(-\frac{1}{\tilde{T}}\right) - \frac{\tilde{b}_g}{1+c_g\tilde{T}} Ei\left(-\frac{1+c_g\tilde{T}}{\tilde{T}}\right)\right) \tag{13}$$

is a dimensionless function of dimensionless variable $\tilde{T}$, with $\tilde{b}_g = b_g \exp(c_g) \approx 0.9249$ and $Ei(z)$ being the exponential integral function.[52] Similarly, the rate coefficient for the impact ionization of an excited-state atom is $\gamma_{ie} = (7/8)\gamma_{ig} + (1/8)\gamma_{iplus}$, with $\gamma_{iplus} = \Gamma_{ex}\tilde{\gamma}_{iplus}$, where $\Gamma_{ex} \approx \Gamma_g/(1-\tilde{\varepsilon}_{ex})$ and

$$\tilde{\gamma}_{iplus} = -\frac{1}{\sqrt{\tilde{T}}}\left(Ei\left(-\frac{1-\tilde{\varepsilon}_{ex}}{\tilde{T}}\right) - \frac{\tilde{b}_{ex}(1-\tilde{\varepsilon}_{ex})}{1-\tilde{\varepsilon}_{ex}+c_{ex}\tilde{T}} Ei\left(-\frac{1-\tilde{\varepsilon}_{ex}+c_{ex}\tilde{T}}{\tilde{T}}\right)\right) \tag{14}$$

With $\tilde{b}_{ex} = b_{ex}\exp(c_{ex})$ and $\tilde{\varepsilon}_{ex} = \varepsilon_{ex}/\varepsilon_{ion}$. The thermal averaging of Eq. (2) results in the impact excitation rate coefficient $\nu_{ex} = \Delta_{ex}\tilde{\nu}_{ex}(\tilde{T})$, where

$\Delta_{ex} = 2.80\pi^2 (Ry/\hbar) a_B^{\,3} (2Ry/(\pi\varepsilon_{ion}))^{3/2}$ and

$$\tilde{v}_{ex} = \frac{1}{\left(\tilde{\varepsilon}_{ex}\right)^{3/2}} \left(\frac{\tilde{\varepsilon}_{ex}}{\tilde{T}}\right)^{3/8} e^{-\frac{\tilde{\varepsilon}_{ex}}{2\tilde{T}}} W_{-15/8,-5/8}\left(\frac{\tilde{\varepsilon}_{ex}}{\tilde{T}}\right), \tag{15}$$

where $W_{\mu,\nu}(z)$ is the Whittaker function.[52]

Further, we use the initial density of neutral atoms, $n_0$, as the concentration scale, and thus the dimensionless densities of ions, excited atoms, and electrons are $\tilde{n}_i = n_i/n_0$, $\tilde{n}_{ex} = n_{ex}/n_0$, and $\tilde{n} = n/n_0$, respectively. Then, the elastic scattering rate is presented as $\nu = \Delta_\nu \tilde{v}(\tilde{T}, \tilde{n}, \tilde{n}_i)$, where $\Delta_\nu = \pi(Ry/\hbar)\sqrt{\alpha}\left(n_0 a_0^3\right)\left(a_B/a_0\right)^{3/2}$ with $a_0 = \sqrt{\sigma_0/\pi} = 1.784 \cdot 10^{-8}$ cm, and $\tilde{v}(\tilde{T}, \tilde{n}, \tilde{n}_i)$ is a function of the dimensionless variables,

$$\tilde{v} = \sqrt{\tilde{T}}\left((1-\tilde{n}_i) + \frac{1}{2\pi\alpha^2}\frac{\tilde{n}_i}{\tilde{T}^2}\ln\left(\frac{\alpha^3}{36\pi n_0 a_0^3}\frac{\tilde{T}^3}{\tilde{n}}\right)\right) \tag{16}$$

In this expression, the first term corresponds to the scattering on neutral atoms, and the second term corresponds to long-range Coulombic scattering on ions.

Then, we choose the formal laser beam radius, $r_0 = 10\ \mu$m, as the scale of the radial coordinate, so that $r = r_0\tilde{r}$. The electron velocity scale $u_0$ comes about as $u_0 = \sqrt{\varepsilon_{ion}/m}$ and gives the time scale, $t_0 = r_0/u_0 = r_0\sqrt{m/\varepsilon_{ion}} \approx 5.998$ ps. The electric field scale, $E_0$, is the field magnitude that provides the electron energy shift of $\varepsilon_{ion}$ over the distance of $r_0$: $E_0 = \varepsilon_{ion}/(er_0) = 1.58 \cdot 10^6$ V/m. In the dimensionless form, the system of equations reads:

$$\frac{\partial \tilde{n}}{\partial \tilde{t}} + \frac{1}{\tilde{r}}\frac{\partial}{\partial \tilde{r}}(\tilde{r}\tilde{n}\tilde{u}) = \lambda\tilde{n}\left(\tilde{\gamma}_{ig}(1-\tilde{n}_i) + \frac{1}{8}(\mu\tilde{\gamma}_{iplus} - \tilde{\gamma}_{ig})\tilde{n}_{ex}\right) + \lambda\tilde{\gamma}_P\ \tilde{n}_{ex}^2 \tag{17}$$

$$\frac{\partial \tilde{u}}{\partial \tilde{t}} = -\frac{1}{\tilde{n}}\frac{\partial}{\partial \tilde{r}}(\tilde{n}\tilde{T}) + \tilde{E} - \lambda\tilde{u}\left(\tilde{\gamma}_{ig}(1-\tilde{n}_i) + \frac{1}{8}(\mu\tilde{\gamma}_{iplus} - \tilde{\gamma}_{ig})\tilde{n}_{ex} + \tilde{\gamma}_P\frac{\tilde{n}_{ex}^2}{\tilde{n}}\right) - \eta\tilde{v}\tilde{u} \tag{18}$$

$$\frac{\partial \tilde{T}}{\partial \tilde{t}} + \tilde{u}\frac{\partial \tilde{T}}{\partial \tilde{r}} + \frac{2}{3}\frac{\tilde{T}}{\tilde{r}}\frac{\partial}{\partial \tilde{r}}(\tilde{r}\tilde{u}) - \frac{1}{\eta}\frac{1}{\tilde{n}\tilde{r}}\frac{\partial}{\partial \tilde{r}}\left(\tilde{r}\frac{\tilde{n}\tilde{T}}{\tilde{v}}\frac{\partial \tilde{T}}{\partial \tilde{r}}\right) = -\lambda\left(\tilde{\gamma}_{ig}\left(\tilde{T}+\frac{2}{3}\right) + \frac{2}{3}\delta\tilde{v}_{ex}\tilde{\varepsilon}_{ex}\right)(1-\tilde{n}_i) - $$

$$\lambda\left(\frac{1}{8}(\mu\tilde{\gamma}_{iplus} - \tilde{\gamma}_{ig})\left(\tilde{T}+\frac{2}{3}\right) - \frac{2}{3}\left(\delta\tilde{v}_{ex} + \frac{7}{8}\tilde{\gamma}_{ig} + \frac{1}{8}\mu\tilde{\gamma}_{iplus}\right)\tilde{\varepsilon}_{ex} - \tilde{\gamma}_P\frac{\tilde{n}_{ex}}{\tilde{n}}\left(\tilde{T}+\frac{2}{3}(1-2\tilde{\varepsilon}_{ex})\right)\right)\tilde{n}_{ex}$$

(19)

$$\frac{1}{\tilde{r}}\frac{\partial}{\partial \tilde{r}}(\tilde{r}\tilde{E}) = -\xi(\tilde{n}-\tilde{n}_i) \tag{20}$$

$$\frac{\partial \tilde{n}_i}{\partial \tilde{t}} = \lambda\tilde{n}\left(\tilde{\gamma}_{ig}(1-\tilde{n}_i) + \frac{1}{8}(\mu\tilde{\gamma}_{iplus} - \tilde{\gamma}_{ig})\tilde{n}_{ex}\right) + \lambda\tilde{\gamma}_P\tilde{n}_{ex}^2 \tag{21}$$

$$\frac{\partial \tilde{n}_{ex}}{\partial \tilde{t}} = \lambda\tilde{n}\left(\delta\tilde{v}_{ex}(1-\tilde{n}_i) - \left(\frac{7}{8}\tilde{\gamma}_{ig} + \frac{1}{8}\mu\tilde{\gamma}_{iplus} + \delta\tilde{v}_{ex}\right)\tilde{n}_{ex}\right) - \lambda\tilde{\gamma}_P\tilde{n}_{ex}^2 \tag{22}$$

with the expressions for the dimensionless constants:

$$\alpha = \frac{\varepsilon_{ion}}{Ry}\frac{a_0}{a_B}; \quad \xi = 4\pi\frac{e^2 r_0^2 n_0}{\varepsilon_{ion}}; \quad \lambda = \Gamma_g n_0 t_0; \quad \eta = \Delta_\nu t_0; \quad \mu = \frac{\tilde{a}_{ex}}{\tilde{a}_{ig}(1-\tilde{\varepsilon}_{ex})}; \quad \delta = \frac{2.80\pi}{\tilde{a}_{ig}}. \tag{23}$$

The numerical values of the constants defined in Eq. (23) are found to be: $\alpha = 1.95$; $\lambda = 2835$; $\eta = 1439.7$; $\xi = 2069\cdot 10^4$; $\delta \approx 0.745$; $\mu = 3.98$.

## 3. Approximations and numerical solution

The fact that the parameter $\xi \gg 1$ allows one to use the following approach, which breaks the wake channel evolution into two distinct stages. First, from Eqs. (21), (17), and (20) it follows that

$$\frac{\partial \tilde{E}}{\partial \tilde{t}} = \xi\tilde{n}\tilde{u}. \tag{24}$$

Then, Eq. (18) is recast as

$$\frac{\partial^2 \tilde{E}}{\partial \tilde{t}^2} + \eta\tilde{v}\frac{\partial \tilde{E}}{\partial \tilde{t}} - \xi\tilde{n}\tilde{E} = -\xi\frac{\partial}{\partial \tilde{r}}(\tilde{n}\tilde{T}) - \frac{1}{\xi}\frac{1}{\tilde{n}\tilde{r}}\frac{\partial \tilde{E}}{\partial \tilde{t}}\frac{\partial}{\partial \tilde{r}}\left(\tilde{r}\frac{\partial \tilde{E}}{\partial \tilde{t}}\right) \tag{25}$$

As $\xi \gg 1$, the last term in the right-hand side of this equation can be neglected, and the first term considered as virtually time-independent. Then, the fast initial evolution of $\tilde{E}$ is easily found as:

$$\tilde{E}(\tilde{t}) \approx \frac{1}{\tilde{n}}\frac{\partial}{\partial \tilde{r}}(\tilde{n}\tilde{T})(1-\exp(-\beta\tilde{t})), \qquad (26)$$

where $\beta = -(\eta\tilde{v}/2) - \sqrt{(\eta\tilde{v}/2)^2 + \xi\tilde{n}}$ being a large parameter justifies the approximation. For $\tilde{t} > 1/\beta$, the function $\tilde{E}(\tilde{t})$ assumes its slower pace, and at the second stage of the evolution,

$$\tilde{E} \approx \frac{1}{\tilde{n}}\frac{\partial}{\partial \tilde{r}}(\tilde{n}\tilde{T}) \approx \frac{1}{\tilde{n}_i}\frac{\partial}{\partial \tilde{r}}(\tilde{n}_i\tilde{T}) \qquad (27)$$

Then, at this slower evolution stage, from Eq. (20) we have

$$\tilde{n} - \tilde{n}_i \approx \frac{1}{\xi\tilde{r}}\frac{\partial}{\partial \tilde{r}}\left(\frac{\tilde{r}}{\tilde{n}_i}\frac{\partial}{\partial \tilde{r}}(\tilde{n}_i\tilde{T})\right) \qquad (28)$$

The physical meaning of Eqs. (27)-(28) is that the electron distribution is slightly wider than the ion distribution, because of the high electron temperature, and this local charge density disbalance induces the radial electric field directed outward from the channel axis. The dimensionless radial velocity of the electron gas is obtained from Eq. (24) as:

$$\tilde{u} = -\frac{1}{\xi\tilde{n}_i}\frac{\partial}{\partial \tilde{t}}\left(\frac{1}{\tilde{n}_i}\frac{\partial}{\partial \tilde{r}}(\tilde{n}_i\tilde{T})\right). \qquad (29)$$

As seen, this average electron velocity has indeed a negligibly small value $\sim 1/\xi$.

We substitute $(\tilde{n} - \tilde{n}_i)$ from Eq. (28) and $\tilde{u}$ from Eq. (29) in Eq. (19) and neglect the terms on the order of $1/\xi^2$. In this way, the equation for the electron temperature is obtained as,

$$\frac{\partial \tilde{T}}{\partial \tilde{t}} = \frac{1}{\xi} \frac{1}{\tilde{n}_i} \frac{1}{\tilde{r}} \frac{\partial}{\partial \tilde{r}} \left( \tilde{r} \frac{\tilde{n}_i \tilde{T}}{\tilde{v}} \frac{\partial \tilde{T}}{\partial \tilde{r}} \right) - \left( \tilde{\gamma}_{ig} \left( \tilde{T} + \frac{2}{3} \right) + \frac{2}{3} \delta \tilde{v}_{ex} \tilde{\varepsilon}_{ex} \right) (1 - \tilde{n}_i) -$$

$$\left( \frac{1}{8} (\mu \tilde{\gamma}_{iplus} - \tilde{\gamma}_{ig}) \left( \tilde{T} + \frac{2}{3} \right) - \frac{2}{3} \left( \delta \tilde{v}_{ex} + \frac{7}{8} \tilde{\gamma}_{ig} + \frac{1}{8} \mu \tilde{\gamma}_{iplus} \right) \tilde{\varepsilon}_{ex} - \tilde{\gamma}_P \frac{\tilde{n}_{ex}}{\tilde{n}_i} \left( \tilde{T} + \frac{2}{3} - \frac{4}{3} \tilde{\varepsilon}_{ex} \right) \right) \tilde{n}_{ex} +$$

(30)

$$\frac{\tilde{T}}{\xi} \left( \frac{1}{\eta \tilde{v}} \frac{\partial \tilde{T}}{\partial \tilde{r}} \frac{\partial}{\partial \tilde{r}} \left( \frac{1}{\tilde{r} \tilde{n}_i} \frac{\partial}{\partial \tilde{r}} \left( \frac{\tilde{r}}{\tilde{n}_i} \frac{\partial}{\partial \tilde{r}} (\tilde{n}_i \tilde{T}) \right) \right) + \frac{1}{\tilde{n}_i} \left( \frac{1}{\tilde{T}} \frac{\partial \tilde{T}}{\partial \tilde{r}} + \frac{2}{3 \tilde{r}} \right) \frac{\partial}{\partial \tilde{t}} \left( \frac{1}{\tilde{n}_i} \frac{\partial}{\partial \tilde{r}} (\tilde{n}_i \tilde{T}) \right) +$$

$$\frac{2}{3} \frac{\partial}{\partial \tilde{r}} \left( \frac{1}{\tilde{n}_i} \frac{\partial}{\partial \tilde{t}} \left( \frac{1}{\tilde{n}_i} \frac{\partial}{\partial \tilde{r}} (\tilde{n}_i \tilde{T}) \right) \right) \right)$$

In this equation, all the terms in the third and the fourth lines are on the order of $1/\xi$ and thus constitute small corrections. Lastly, we explicitly separate the $1/\xi$-order corrections in the equations for the local concentrations of the ions and the excited atoms:

$$\frac{\partial \tilde{n}_i}{\partial \tilde{t}} \approx \lambda \tilde{n}_i \left( \tilde{\gamma}_{ig} (1 - \tilde{n}_i) + \frac{1}{8} (\mu \tilde{\gamma}_{iplus} - \tilde{\gamma}_{ig}) \tilde{n}_{ex} \right) + \lambda \tilde{\gamma}_P \tilde{n}_{ex}^2 +$$

(31)

$$\frac{\lambda}{\xi} \left( \tilde{\gamma}_{ig} (1 - \tilde{n}_i) + \frac{1}{8} (\mu \tilde{\gamma}_{iplus} - \tilde{\gamma}_{ig}) \tilde{n}_{ex} \right) \frac{1}{\tilde{r}} \frac{\partial}{\partial \tilde{r}} \left( \frac{\tilde{r}}{\tilde{n}_i} \frac{\partial}{\partial \tilde{r}} (\tilde{n}_i \tilde{T}) \right)$$

and

$$\frac{\partial \tilde{n}_{ex}}{\partial \tilde{t}} \approx \lambda \tilde{n}_i \left( \delta \tilde{v}_{ex} (1 - \tilde{n}_i) - \left( \frac{7}{8} \tilde{\gamma}_{ig} + \frac{1}{8} \mu \tilde{\gamma}_{iplus} + \delta \tilde{v}_{ex} \right) \tilde{n}_{ex} \right) - \lambda \tilde{\gamma}_P \tilde{n}_{ex}^2 +$$

(32)

$$\frac{\lambda}{\xi} \left( \delta \tilde{v}_{ex} (1 - \tilde{n}_i) - \left( \frac{7}{8} \tilde{\gamma}_{ig} + \frac{1}{8} \mu \tilde{\gamma}_{iplus} + \delta \tilde{v}_{ex} \right) \tilde{n}_{ex} \right) \frac{1}{\tilde{r}} \frac{\partial}{\partial \tilde{r}} \left( \frac{\tilde{r}}{\tilde{n}_i} \frac{\partial}{\partial \tilde{r}} (\tilde{n}_i \tilde{T}) \right)$$

Thus, in the first approximation, the evolving distributions of the ion concentration, the excited-atom concentration, and the electron temperature are determined by the system of coupled "diffusion-reaction" equations, which consists of the first line in Eq. (31), the first line in Eq. (32), and the first two lines in Eq. (30). In this approximation, the hydrodynamic velocity of the electron gas is negligible, and the radial electric field is expressed by Eq. (27). The corrections on the order of $1/\xi$ can be found by substituting the obtained $T(r,t)$, $n_i(r,t)$, and $n_{ex}(r,t)$ in the remaining terms in Eqs. (30), (31), and (32), as well as in the right-hand sides of Eqs. (28) and (29), and using these as the source terms.

As was mentioned earlier, the initial distributions of $T(r)$, $n_i(r)$, and $n_{ex}(r)$ at the onset of the wake channel evolution are obtained by solving Eq. (1) for the build-up of inhomogeneous excited state of the dense-gas medium during the laser pulse. The combination of partial-derivative terms, finite-difference terms, and an integral term in the right-hand side of this equation requires specially tailored *ad hoc* approach to numerical solution of this equation.[40] We have implemented an approach similar to that developed previously for delayed differential equations.[53] To mimic typical experimental conditions,[16] the dense-gas medium is assumed to be argon at the pressure of 60 atm and room temperature. The laser pulse of $7 \times 10^{13}$ W/cm² intensity and 800 nm carrier wavelength is modeled by the envelope function with the radial Gaussian profile and cosine-squared temporal profile, $s(r,t) = \cos^2\left((\pi/\tau)(t-\tau/2)\right)\Theta(t)\Theta(t-\tau)\exp(-r^2/r_0^2)$, with $\tau = 80$ fs, which corresponds to the pulse FWHM of 40 fs, and with the beam radius of $r_0 = 10$ µm. We use the ADK formula[54] for the strong-field ionization rate $W(t)$, and solve Eq. (1) numerically for the evolution of the local electron energy distribution, $n(r,\varepsilon,t)$ and the corresponding density distributions of the ions, $n_i(r,t)$ and the excited atoms, $n_{ex}(r,t)$, during the laser pulse. The distribution of average electron energy is then found as $\langle \varepsilon \rangle = \int_0^\infty d\varepsilon\, \varepsilon\, n(\varepsilon,r,t) \Big/ \int_0^\infty d\varepsilon\, n(\varepsilon,r,t)$, determining the radial profile of electron temperature, $T(r) = (2/3)\langle\varepsilon\rangle$, which is established in the immediate wake of the pulse.

Fig. 2 shows the radial profiles of the ion density, $n_i(r)$, and the density of excited atoms, $n_{ex}(r)$, after the end of the laser pulse. The inset shows how these densities have been evolving during the laser pulse on the axis of the filament channel. As seen in the inset, the initial surge of ionic density near the middle of the pulse is soon overcome by the steady growth of the excited-atom density, so that by the end of the pulse the concentration of the excited atoms at the channel axis is about three times greater than the concentration of ions. This disparity in the final concentrations persists across the channel, although the $n_{ex}/n_{ion}$ ratio becomes smaller away from the channel axis. This prevalence of the excitation over ionization reflects the differences in the dependences of the respective cross-sections on the electron energy as discussed in the previous section.

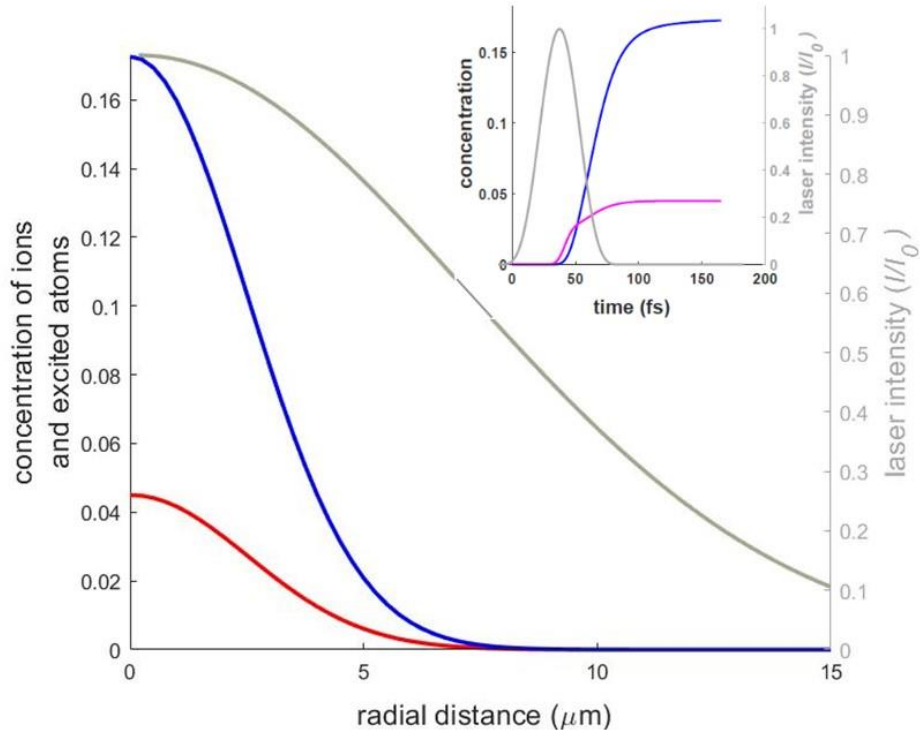

**Figure 2**. Radial concentration profiles of the ionized (red curve) and excited (blue curve) atoms in the filament channel in the immediate pulse wake. The beam intensity profile is shown as the gray line for comparison. Inset: evolution of the ionic and excited-atom concentrations on the channel axes during the laser pulse, whose envelope is shown as the grey line.

Fig. 3 shows the radial profile of the effective electron temperature, $T(r)$, after the pulse, and the inset shows the evolution of the average electron energy during the pulse at the channel axis. As seen in the inset, initially the average energy grows on the order of 20 fs, being effectively supplied by the inverse Bremsstrahlung. However, as the laser pulse subsides, the average energy decreases considerably, reflecting the energy consumption by the impact excitation and ionization processes. The resulting electron temperature profile is wider than that of the laser pulse. However, when compared to the profiles in Fig. 2, the wider region corresponds to low electron density and thus contains very small amount of excess energy to be utilized in the after-pulse evolution of the system.

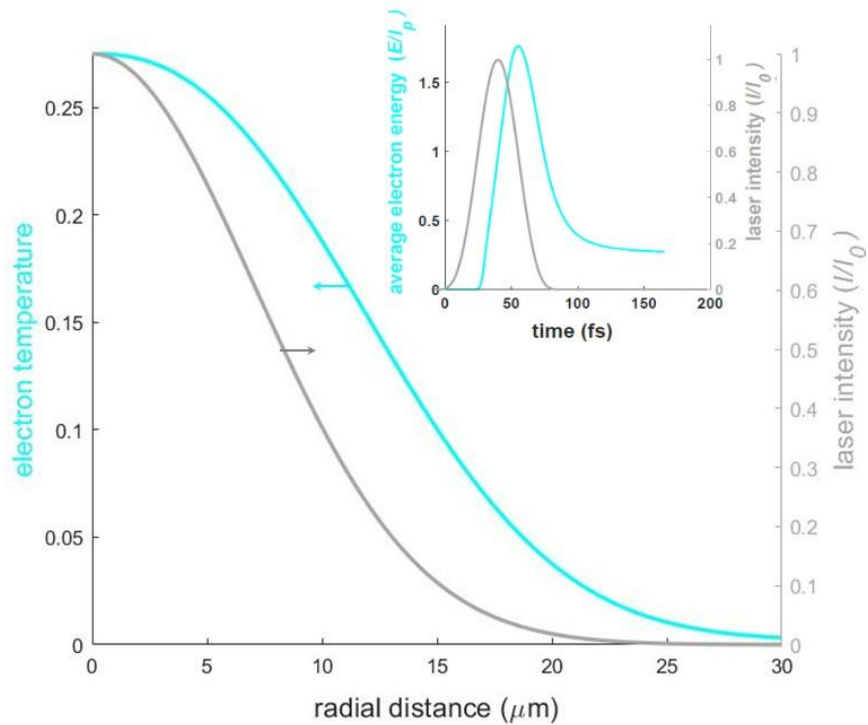

**Figure 3**. Radial profile of the electron temperature (in the units of $I_p$) upon thermalization in the immediate pulse wake. The radial intensity profile of the laser beam is shown for comparison as the gray line. The inset shows evolution of the average electron energy during the pulse.

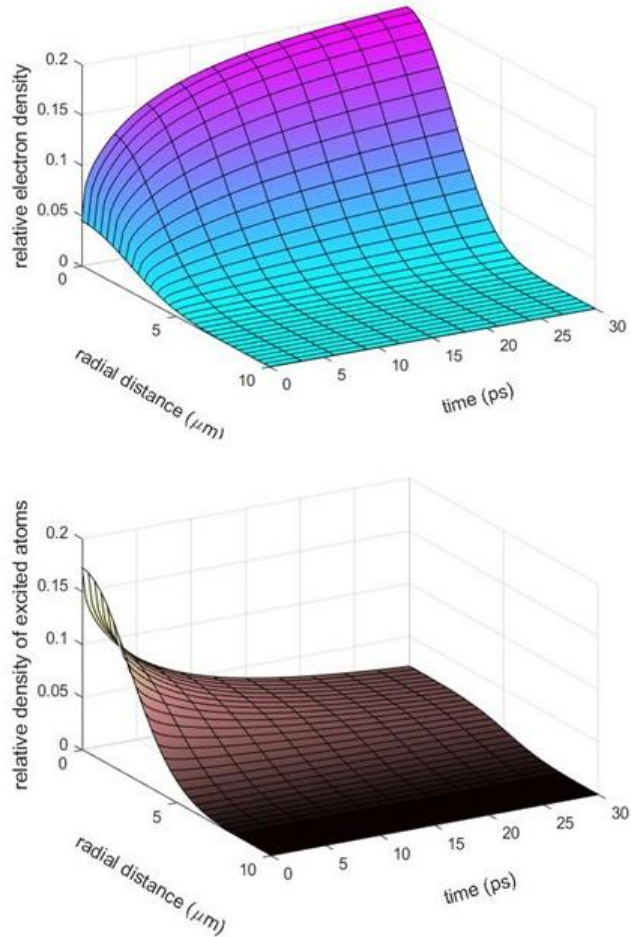

**Figure 4**. Evolution patterns of the ionic concentration (upper panel) and the excited atoms concentration in the filament wake channel in argon at 60 atm pressure, upon interaction with the laser pulse of $7\times10^{13}$ W/cm$^2$ intensity.

The obtained distributions $n_{ex}(r)$, $n_{ion}(r)$, and $T(r)$ set the stage for the filament channel evolution in the wake of the laser pulse. We explored this evolution by numerically solving the system of equations (30), (31), (32) and then reconstructing the electric field profile via Eq. (27). The resulting patterns of the concerted evolution of the ionic density and the excited-atom density are shown in Fig. 4. As seen in the Figure, the evolution of the excited-atom density mirrors the evolution of the ion density, so during the entire process the increase in the ionic density is coming mostly at the expense of the decreasing density of excited atoms. Thus, a supply of new ions in the channel is mediated by excited atoms. Further, the evolution of both densities involves two distinct stages: the initial fast evolution during about 3 ps is followed by a slower process taking about 30 ps.

The ionization dynamics of the wake channel is reflected in its linear and nonlinear optical characteristics. In particular, the grows of the free-electron density leads to increase in the negative contribution to the refractive index of the medium, which directly manifests itself in the phase shift of the probe beam upon crossing the channel, as was observed in Ref. 16. It is instructive to compare the experimentally observed phase shifts with those predicted on the basis of our model analysis. In Ref. 16, the phase shift was measured at three values of the argon gas pressure: 60 bar, 40 bar and 20 bar, and the parameters of the laser filament varied accordingly, due to the variation of the critical power for self-focusing. Using these parameters in our model numerical simulations, we have obtained the time-dependent electron concentration values $\tilde{n}(r,t)$ in these three cases, and calculated the expected phase shifts using the approximate expression, $\Delta\varphi(t) = (k n_0 / N_c) \int_0^\infty dr\, \tilde{n}(r,t)$, where $k = 2\pi/\lambda$ is the laser angular wavenumber and $N_c = \pi m_e c^2 / (e^2 \lambda^2)$ is the critical plasma density, $\lambda$ being the laser wavelength. These phase shifts as functions of the pump-probe delay time are presented in Fig. 5. As seen, they agree reasonably well with the experimental measurements reported in Fig. 2 (b) of Ref. 16, which is reproduces in the inset. One can notice, however, that the theoretical curves demonstrate a more rapid increase in the beginning of the wake channel evolution, as compared to the experimental ones. This difference may be related to the appreciable concentration of the excited atoms in the immediate wake of the laser pulse. The polarizability of these excited atoms is considerably larger than that of the neutral atoms, providing a positive contribution to the refractive index and thus weakening the effect of the initial electron concentration.

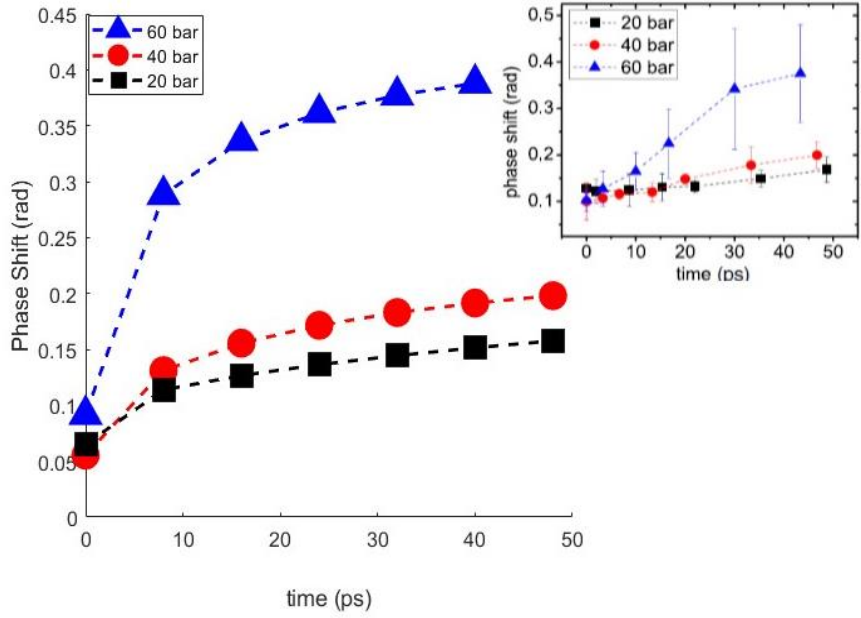

**Figure 5**. Simulated phase shift of the pump pulse crossing the wake channel as a function of the pump-probe delay. The filament characteristics are those in Ref. 16. The inset: experimental measurements of the phase shifts reported in Fig. 2 of Ref. 16.

The cause of the two-scale behavior can be seen in Figure 6, where a similar behavior is demonstrated by the temperature distribution. As the coefficients of the impact processes in Eqs. (13)-(16) do depend appreciably on temperature, initially, when the temperature is high, the processes go fast and they rapidly consume the energy of the electron gas. The longer process is sustained by ongoing Penning ionization which tends to increase the temperature when releasing the excess energy of two excited atoms into the kinetic energy of the emerging electron. The slower evolution process is thus driven by the tail of electron thermal distribution which is sustained by the Penning ionization. Another interesting feature of the electron temperature evolution presented in Figure 6 is that the temperature distribution develops and maintains a non-monotonic radial profile, so that the temperature peaks not at the channel axes but at the periphery.

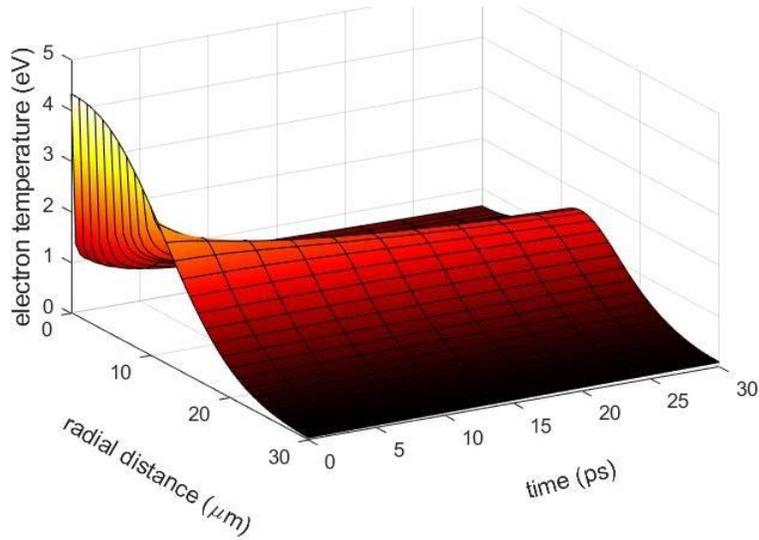

**Figure 6**. The evolution of the electron temperature distribution in the filament wake channel in argon at 60 atm pressure, upon interaction with the laser pulse of $7\times10^{13}$ W/cm$^2$ intensity.

The radial electric field that develops in the channels during the electronic evolution is caused by a local imbalance of electron and ion densities and is given in the main approximation by Eq. (27), thus depending both on the ion density distribution and the temperature distribution in the channel. The evolution of this electric field is presented in

Figure 7. As seen, the mentioned nonmonotonic radial temperature profile leads to a nonmonotonic radial profile of the electric field with the peak values on the order of $10^7$ V/m. This considerable built-in electric field is supposed to factor in the nonlinear optical response of the channel.

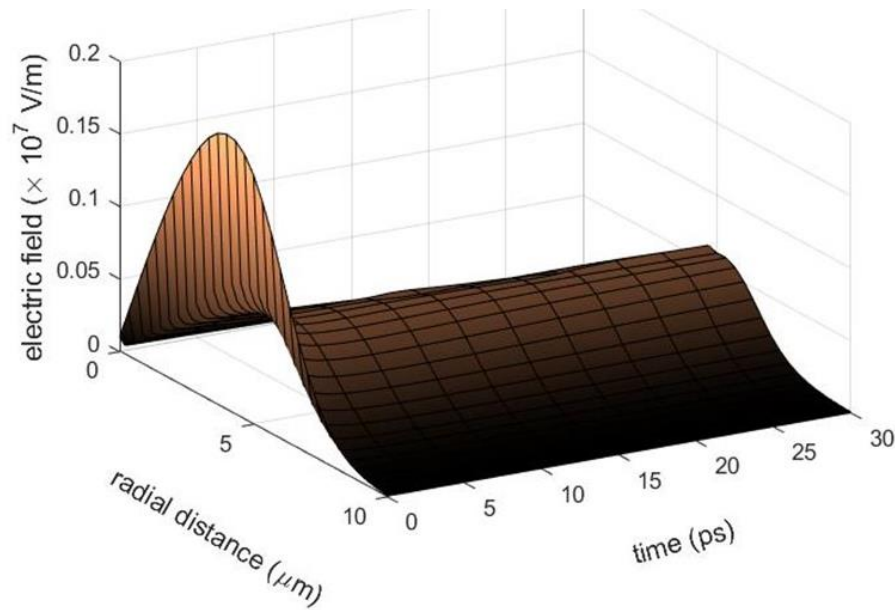

**Figure 7**. The evolution of the radial electric field in the filament wake channel in argon at 60 atm pressure, upon interaction with the laser pulse of $7\times10^{13}$ W/cm$^2$ intensity.

## 4. Conclusions

We have developed a unified theoretical description of the formation of an ionized filament channel in a dense-gas medium and the electronic evolution of this channel in the laser pulse wake. The processes involved are illustrated on the experimentally important example of filaments produced by intense, femtosecond 800-nm laser pulses in high-pressure argon gas.

Unlike the situation with atmospheric-pressure gases, the ionization and excitation dynamics during the laser pulse are mainly driven by the inverse Bremsstrahlung of the emerging free electrons on neutral atoms. This process provides the energy gain that enables collisional excitation and impact ionization processes. A kinetic model of these interdependent processes allows one to obtain the radial density distributions in the immediate wake of the laser pulse and predicts the prevalence of excited atoms over ionized atoms.

In the wake of the laser pulse, the thermalized electron gas drives the system evolution via the impact ionization (from the ground and from the excited states) and collisional excitation of the residual ground-state neutral atoms, which processes are affected by the thermal conduction in the gradually cooling electron gas. The massively present and continually supplied excited atoms are actively engaged in the process of Penning ionization, thus slowing the electron gas cooling. During this evolution of the electronic degrees of freedom, the local imbalance of the free-electron and ion densities creates and maintains a transient radial electric field on the order of $10^7$ V/m.

For the case of argon at 60 atm pressure interacting with a laser pulse of $7 \times 10^{13}$ W/cm$^2$ intensity, we have solved numerically the Fokker-Planck type equations that describe the energy intake and redistribution during the pulse, when the electrons are released via strong-field ionization and forcefully driven by the oscillating laser field, colliding mainly with neighboring neutral atoms. The outcome radial density distributions of the free electrons, ions, and excited neutral atoms at the end of the laser pulse show considerable (about three-fold) prevalence of excited atoms over ionized atoms. The obtained radial profile of the average electron energy provides the initial electron temperature distribution upon thermalization in the immediate wake of the pulse.

The evolution of electronic degrees of freedom in the filament wake channel is described by a system of coupled diffusion-reaction equations for the electron density, the ion density, the density of excited atoms, the electron temperature, the electron gas velocity, and the induced radial electric field. Solving numerically these equations, we have obtained the evolving radial profiles of all these characteristics of the wake channel. An important revealed feature of the wake-channel evolution is that it proceeds with two characteristic temporal scales. First, the fast evolution takes place within ~ 5 ps immediately after the laser pulse. This fast rate is determined by the initial high concentration of excited atoms and high electron temperature, and the evolution consists in fast cooling of the free electron gas accompanied by extensive ionization of the excited atoms. Then, the impact processes assume slower rate as determined by the lowered electron temperature, which it turn is sustained by the electron kinetic energy release in the ongoing Penning ionization; this slower evolution stage takes about 30 ps. The obtained spatiotemporal dependences of electron and excited-atom densities and the built-in radial electric field should allow for experimental tracing of the wake channel evolution via linear and nonlinear light-scattering experiments.

## 5. Acknowledgments

This work was supported by the National Science Foundation under Grant No. PHY1806594 and by the Office of Naval Research under Award No. N00014-15-1-2574.